\documentclass[conference]{IEEEtran}

\usepackage{amsmath,amssymb,amsfonts,amsthm,mathtools}

\usepackage[utf8]{inputenc}
\usepackage[T1]{fontenc}
\usepackage{url}
\usepackage{booktabs}
\usepackage{nicefrac}
\usepackage{microtype}
\usepackage{xcolor}
\usepackage{graphicx}
\usepackage{tikz}
\usetikzlibrary{positioning,fit,backgrounds,arrows.meta}
\usepackage{multirow}
\usepackage{cite}
\usepackage[capitalize,noabbrev]{cleveref}

\newcommand{\Vbar}{\bar{V}}       
\newcommand{\Ybus}{\mathbf{Y}}    

\newcommand{\kth}{^{(k)}}         
\newcommand{\knext}{^{(k+1)}}     

\newcommand{\MAE}{\mathrm{MAE}}
\newcommand{\erel}{\varepsilon_{\mathrm{rel}}}  

\newcommand{\pg}{P_g}   
\newcommand{\qg}{Q_g}   

\title{A Benchmark on LLM-Based Power Flow Computation: Do More Structured Prompts Help?
}

\author{
  \IEEEauthorblockN{Tingwei Chen, Kaiyang Huang,~\textit{Student Member, IEEE},
    and Kai Sun,~\textit{Fellow, IEEE}}
  \IEEEauthorblockA{University of Tennessee, Knoxville}
}

\begin{document}

\maketitle

\begin{abstract}
We present a controlled benchmark evaluating three LLMs---Claude Sonnet~4.5,
Gemini~2.5 Pro, and GPT-3.5 Turbo---across four prompt formats (from concise
narrative to structured JSON with explicit iteration trace) on Gauss--Seidel
AC power flow computation for a three-bus system.
Against 50 test cases with reference solutions computed numerically,
Gemini~2.5 Pro with the simplest narrative prompt achieves the lowest mean absolute
error (MAE\,=\,0.257~MW/MVar, 54\% of cases within 5\% relative error), while the
same model with a JSON-structured prompt raises MAE to 0.789---a 3.1$\times$ increase.
Adding a worked example degrades accuracy for Gemini but provides a marginal gain
for Claude.
GPT-3.5 Turbo fails on at least 90\% of cases under all prompt formats.
An independent 100-case replication with related prompt-format families confirms the
qualitative ordering (Gemini\,$\gg$\,Claude\,$>$\,GPT-3.5): the best 100-case
configuration (Gemini with explicit iteration trace) achieves MAE\,=\,0.402 and 53\%
within 5\%, while Claude Sonnet~4.5's near-flat accuracy profile
($\approx$38\% within 5\% across formats) and GPT-3.5's near-total ineffectiveness
(92--97\% above 20\% error) both replicate.
In neither evaluation does any configuration achieve sufficient reliability for use
as a direct numerical solver.
These findings offer a diagnostic baseline for practitioners and researchers
evaluating LLMs for smart-grid decision-support assistance.

\end{abstract}


\section{Introduction}
\label{sec:intro}

Power flow computation is the computational foundation of grid planning, operation,
and stability analysis.
As smart-grid deployments grow increasingly complex, operators and engineers rely
on decision-support tools that can perform or explain numerical computations rapidly.
Large language models (LLMs) are attractive candidates for such copilot roles---from
explaining solver output to assisting with engineering feasibility checks.
The Gauss--Seidel method, introduced decades ago and still taught as the canonical
iterative solver in power engineering curricula~\cite{saadat2010power}, requires
careful execution of complex arithmetic across multiple convergence iterations.
As LLMs attract growing interest for power system tasks~\cite{ruan2024security},
a natural question arises: can these models reliably execute a classical iterative
numerical algorithm---and does the way we ask matter?

Prior LLM-for-power-systems work has focused on high-level tasks such as fault diagnosis,
maintenance Q\&A, and security analysis~\cite{ruan2024security}.
These tasks rely on text understanding and knowledge retrieval rather than precise
arithmetic.
Separately, the prompt engineering literature has shown that chain-of-thought
decomposition and few-shot examples can improve LLM performance on mathematical
reasoning~\cite{wei2022chain,brown2020language}, but these benchmarks typically
involve closed-form problems.
Gauss--Seidel AC power flow requires multi-pass convergence, complex number arithmetic,
and careful tracking of intermediate voltage estimates across iterations---a qualitatively
different challenge.

No prior study has systematically varied prompt representation for a classical iterative
AC power flow solver in a controlled setting.
We close this gap with a benchmark that systematically varies prompt format while
holding the algorithm, test network, and evaluation protocol constant, enabling a
diagnostic comparison of four prompt-package designs.

Our paper makes three contributions:
\begin{enumerate}
  \item We design four prompt formats spanning narrative, example-augmented, procedural-text,
    and JSON-structured representations for LLM-based Gauss--Seidel power flow.
  \item We show that Gemini~2.5 Pro's overall MAE varies by a factor of 3.1$\times$
    across prompt formats---with the concise narrative format yielding the lowest
    error and the most complex format (structured JSON input with required iteration
    trace output) performing worst.
  \item We show that GPT-3.5 Turbo places $\ge$90\% of cases above 20\% relative error
    regardless of prompt design, and that no configuration achieves reliable accuracy
    sufficient for direct use as a numerical solver.
\end{enumerate}

\begin{figure}[t]
  \centering
  \includegraphics[width=\columnwidth]{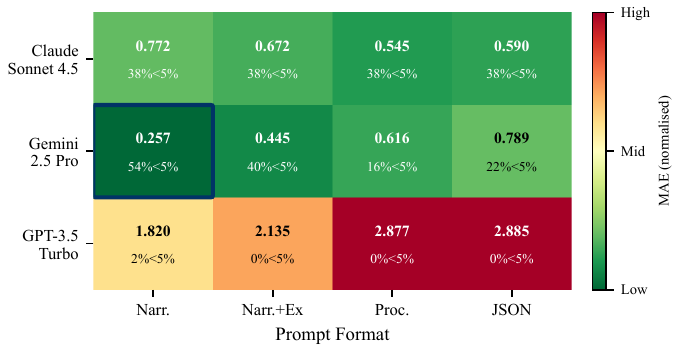}
  \caption{MAE heatmap for 12 LLM--prompt configurations on 50 three-bus
    Gauss--Seidel power flow test cases.}
  \label{fig:mae_heatmap}
\end{figure}

The best result in our benchmark---Gemini~2.5 Pro with a narrative prompt---achieves
overall MAE = 0.257~MW/MVar, with 54\% of cases within 5\% relative error
(Fig.~\ref{fig:mae_heatmap}; darker green = lower MAE, each cell reports the
fraction of cases with relative error $<$5\%, bold border marks the best entry).
The same model with a JSON-structured prompt yields MAE = 0.789.
This result challenges a common assumption in prompt engineering---that providing
more explicit numerical instructions always improves accuracy on arithmetic tasks.

The remainder of this paper is organized as follows.
Section~\ref{sec:related} reviews related work.
Section~\ref{sec:method} defines the problem and experimental setup.
Section~\ref{sec:results} presents results and analysis.
Section~\ref{sec:conclusion} concludes.

\section{Related Work}
\label{sec:related}

\subsection{LLMs for power system tasks.}
The application of LLMs to power engineering is an active and recent research area.
Ruan et al.\ \cite{ruan2024security} examine security implications of deploying LLMs
in power system contexts and document how LLMs can be prompted to assist with
grid-related tasks---but treat the LLM as a text interface, not a numerical solver.
More broadly, proposals for LLM-assisted fault diagnosis, load forecasting explanation,
and operator Q\&A all rely on retrieval or classification rather than arithmetic
computation.
To the best of our knowledge, no prior work has benchmarked LLM performance on
direct execution of a classical iterative power flow algorithm under controlled prompt
variation.

\subsection{Prompt engineering for mathematical reasoning.}
Chain-of-thought prompting~\cite{wei2022chain} showed that asking LLMs to produce
explicit intermediate reasoning steps substantially improves accuracy on arithmetic
and commonsense benchmarks.
The zero-shot variant~\cite{kojima2022large} demonstrated that the instruction
``Let's think step by step'' transfers across tasks without labeled examples.
Few-shot prompting~\cite{brown2020language} established that providing in-context
worked examples helps LLMs generalise to structurally similar problems.
However, these benchmarks use problems with closed-form or single-step solutions.
Gauss--Seidel iteration requires the model to maintain a convergence trajectory across
multiple passes, updating complex voltage phasors at each step.
Whether chain-of-thought-style procedural instructions---or worked examples---transfer
to multi-pass iterative algorithms is the question our benchmark directly addresses.

\subsection{Benchmarking LLMs on scientific and engineering tasks.}
Lewkowycz et al.~\cite{lewkowycz2022solving} (Minerva) tested LLMs on quantitative
science problems and found that larger models improved substantially on multi-step
mathematical derivations.
Chen et al.~\cite{chen2021codex} evaluated LLMs on code generation (HumanEval),
showing that models can write executable numerical code.
More recently, program-aided approaches~\cite{gao2023pal,chen2023pot} offload
arithmetic to an interpreter, achieving stronger numerical accuracy than direct
text-based computation.
These approaches contrast with our setup, in which LLMs must carry out
multi-pass arithmetic without tool assistance---the failure mode our benchmark
is designed to expose.
No benchmark to date targets direct LLM execution of power flow iteration,
making our three-model, four-format study the first controlled experiment of
this kind.

\section{Problem Formulation and Experimental Setup}
\label{sec:method}

\subsection{Three-Bus Power System and Reference Solver}
\label{sec:method-system}

We use a three-bus system with one slack bus (bus~1) and two PQ load buses (buses~2
and~3), fully connected by three transmission lines.
Bus~1 is held at $V_1 = 1.05\angle 0^\circ$~pu on a 100-MVA base.
All shunt susceptances are neglected.
Across all 50 test cases, the resistance and reactance of each branch are drawn
independently and uniformly: $R \in [0.010,\,0.015]$~pu and
$X \in [0.050,\,0.070]$~pu (per-unit, 100-MVA base), as summarised in
Table~\ref{tab:system_params}.

\begin{table}[t]
  \centering
  \caption{Branch parameter sampling ranges for the three-bus benchmark system
    (per-unit, 100-MVA base). Each branch is sampled independently and uniformly
    within the stated interval.}
  \label{tab:system_params}
  \begin{tabular}{lcc}
    \toprule
    Line (from--to) & $R$ (pu) & $X$ (pu) \\
    \midrule
    Bus 1--Bus 2 & $[0.010,\ 0.015]$ & $[0.050,\ 0.070]$ \\
    Bus 1--Bus 3 & $[0.010,\ 0.015]$ & $[0.050,\ 0.070]$ \\
    Bus 2--Bus 3 & $[0.010,\ 0.015]$ & $[0.050,\ 0.070]$ \\
    \bottomrule
  \end{tabular}
\end{table}

Across the 50 test cases, load values at buses~2 and~3 are randomised independently
over physically plausible ranges (real power: 200--250~MW per bus; reactive power:
100--130~MVar per bus).
The primary benchmark target is the slack-bus generation: real power $\pg$ (MW) and
reactive power $\qg$ (MVar) at bus~1.
These quantities are chosen because they are the result of full convergence and
represent the final output of the power flow; intermediate voltage magnitudes and
angles are not evaluated here.

Gauss--Seidel was chosen over Newton--Raphson for three reasons.
First, it is the canonical iterative solver in power engineering
curricula~\cite{saadat2010power}, giving LLMs trained on textbook material prior
exposure to the algorithm and making it a fair test of whether models can execute
a method they have plausibly encountered.
Second, each G-S update step reduces to scalar complex-number arithmetic through a
single per-bus formula; Newton--Raphson requires constructing and factorising a
Jacobian matrix at every iteration---a qualitatively harder computational demand
that would confound matrix-arithmetic difficulty with prompt-format effects.
Third, a method where some model--format combinations can partially succeed
preserves the diagnostic value of the benchmark: a Newton--Raphson baseline would
likely produce near-universal failure across all configurations, masking any
prompt-format signal.

The Gauss--Seidel update rule for bus $i$ is:
\begin{equation}
  \Vbar_i\knext
  = \frac{1}{Y_{ii}}
    \left[
      \frac{P_i - jQ_i}{\left(\Vbar_i\kth\right)^*}
      - \sum_{j \neq i} Y_{ij}\,\Vbar_j
    \right],
  \label{eq:gs_update}
\end{equation}
where $Y_{ii}$ and $Y_{ij}$ are elements of the bus admittance matrix $\Ybus$,
and $P_i$, $Q_i$ denote the net injected real and reactive power at bus~$i$.
In Eq.~(\ref{eq:gs_update}), voltages $\Vbar_j$ for $j < i$ are the already-updated
values from the current iteration, while those for $j > i$ are from the previous
iteration---the defining characteristic of Gauss--Seidel (vs.\ Gauss--Jacobi).
Buses are updated sequentially (bus~2 before bus~3) using the most recently
available voltages.
Iterations continue until
\begin{equation}
  \max\!\left(|\Delta V_2|,\,|\Delta\delta_2|,\,|\Delta V_3|,\,|\Delta\delta_3|\right)
  < 10^{-4},
  \label{eq:conv}
\end{equation}
where $\Delta V_i = |V_i^{(k+1)}| - |V_i^{(k)}|$ and $\Delta\delta_i = \angle V_i^{(k+1)} - \angle V_i^{(k)}$ (in radians), with divergence declared after 50 iterations.
Post-convergence, the slack-bus generation is recovered as
$\bar{S}_1 = \bar{V}_1 \cdot \overline{\sum_j Y_{1j}\bar{V}_j}$.

We generated 50 test cases by randomizing load values uniformly over physically
plausible ranges (real power: 100--400~MW per bus; reactive power: 50--200~MVar
per bus) while keeping the network topology fixed.
Of these, 29~cases request the real power $\pg$ and 21~cases request the reactive
power $\qg$ at bus~1 (this split reflects how the original evaluation was structured).
All reference solutions are computed in Python using NumPy with double-precision
arithmetic and verified to converge within the tolerance above.
LLM outputs were parsed by extracting the first floating-point number that appears
after the expected keyword (\textit{Real power} or \textit{Reactive power}) in
each response; outputs where no valid number was found are recorded as failures
and excluded from MAE computation but included in the $>$20\% bin.

To assess result stability, the benchmark was independently replicated on
100~freshly sampled test cases (56~$\pg$ cases, 44~$\qg$ cases) drawn from the
same parameter ranges under the same convergence criteria.
The replication applies the same four prompt-format families
(\textit{narr.-final}, \textit{narr.+example}, \textit{proc.-text},
\textit{struct.-json}) to the same three models; results are reported in
Section~\ref{sec:results-scale}.

\subsection{Experimental Data Pipeline}
\label{sec:method-pipeline}

\begin{figure}[t]
  \centering
  \begin{tikzpicture}[
  scale=0.9, every node/.style={transform shape},
  node distance=0.45cm and 0.35cm,
  box/.style={
    draw, rounded corners=4pt,
    minimum width=2.2cm, minimum height=0.85cm,
    align=center, font=\small,
    fill=white, thick
  },
  arr/.style={->, thick, >=stealth},
  frame/.style={draw, rounded corners=6pt, thick, inner sep=0.38cm}
]

\node[box] (gen)   {Generate Random\\Numerical Values};
\node[box, right=of gen]  (solve) {Solve AC\\Power Flow};
\node[box, right=of solve] (keep) {Only Keep\\Convergent Cases};

\node[box, below=of keep] (query) {Query\\LLM API};
\node[box, left=of query]  (coll)  {Collect\\Output};
\node[box, left=of coll]   (anal)  {Data\\Analysis};

\draw[arr] (gen)   -- (solve);
\draw[arr] (solve) -- (keep);

\draw[arr] (keep)  -- (query);

\draw[arr] (query) -- (coll);
\draw[arr] (coll)  -- (anal);

\begin{scope}[on background layer]
  \node[frame, fit=(gen)(solve)(keep)(query)(coll)(anal),
        label={[font=\bfseries\small, anchor=south west]south west:Python}] {};
\end{scope}

\end{tikzpicture}
  \caption{End-to-end experimental pipeline (all stages implemented in Python).}
  \label{fig:pipeline}
\end{figure}
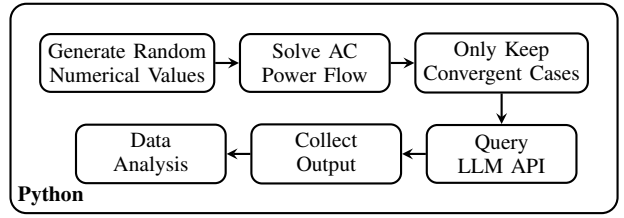

Figure~\ref{fig:pipeline} shows the end-to-end pipeline used to generate test cases,
query the models, and analyse results---all implemented in Python.
The pipeline begins by sampling branch parameters and load values uniformly at random
within the ranges described in Section~\ref{sec:method-system}.
Each sampled configuration is passed to the Gauss--Seidel reference solver; cases that
fail to converge within the prescribed tolerance (Eq.~\ref{eq:conv}) are discarded so
that every benchmark instance has a well-defined ground-truth answer.
The surviving cases are formatted into one of the four prompt styles
(Section~\ref{sec:method-prompts}) and submitted to each LLM API in an independent,
stateless session.
Raw model responses are collected and the target value extracted via keyword-anchored
regex parsing; responses that yield no valid floating-point number are flagged as
failures and handled as described in Section~\ref{sec:method-metrics}.
In the final stage, parsed outputs are compared against the reference solutions to
compute relative errors, MAE, and the error-bin distributions reported in
Section~\ref{sec:results}.

\subsection{LLMs Evaluated}
\label{sec:method-llms}

We benchmark three models representing distinct capability tiers:
(1) \textbf{Gemini~2.5 Pro} (\texttt{gemini-2.5-pro-preview-03-25}, Google DeepMind),
(2) \textbf{Claude Sonnet~4.5} (\texttt{claude-sonnet-4-5}, Anthropic), and
(3) \textbf{GPT-3.5 Turbo} (\texttt{gpt-3.5-turbo}, OpenAI) as a lower-capability baseline.
All queries used the provider default temperature (1.0 for all three APIs) and
the default output-length limit for each provider.
Each case was queried in a fresh session; no retry was performed on parseable outputs.
Experiments were conducted in January 2026; results are snapshot-specific and
may not generalise to later model versions.
Because the default temperature does not guarantee deterministic outputs across all provider
APIs, single-run evaluation is an acknowledged limitation of this study.
The exact prompt texts for all four formats are included in the supplementary material.

\subsection{Prompt Formats}
\label{sec:method-prompts}

We designed four prompt formats of increasing representational complexity
(Table~\ref{tab:prompt_formats}).

\begin{table}[t]
  \centering
  \caption{The four prompt formats evaluated in this study (see text for details).}
  \label{tab:prompt_formats}
  \begin{tabular}{p{1.6cm}p{1.5cm}p{1.1cm}p{2.6cm}}
    \toprule
    Format & Input style & Output & Key design choice \\
    \midrule
    \textit{narr.-final}
      & Natural language
      & Final only
      & Textbook-style question phrasing \\
    \textit{narr.+example}
      & Natural language
      & Final only
      & Adds one solved textbook worked example \\
    \textit{proc.-text}
      & Bullet-point steps
      & Iter.+final
      & Numbered procedure; plain-text data \\
    \textit{struct.-json}
      & JSON object
      & Iter.+final
      & Machine-readable input; written procedure \\
    \bottomrule
  \end{tabular}
\end{table}

\textit{Narrative-final} (\texttt{narr.-final}) presents the network data as a
natural-language paragraph---the same style used in Saadat's textbook~\cite{saadat2010power}---and
asks the model to report only the final power value.
\textit{Narrative+example} (\texttt{narr.+example}) adds one complete worked example
from the same textbook before the target question.
\textit{Procedural-text} (\texttt{proc.-text}) replaces narrative prose with a numbered
five-step Gauss--Seidel procedure and bullet-point numerical data; the model is asked
to output a per-iteration convergence table (BLOCK~1) followed by the final answer
(BLOCK~2).
\textit{Structured-JSON} (\texttt{struct.-json}) encodes all bus and branch data as a
JSON object alongside the written procedure, with the same two-block output requirement.

The formats differ along two dimensions: (i)~input representation
(narrative vs.\ structured) and (ii)~output complexity (final-only vs.\
iteration table).
\texttt{narr.-final} and \texttt{narr.+example} request only the final answer,
while \texttt{proc.-text} and \texttt{struct.-json} additionally require a
full iteration trace.
Complete verbatim prompt texts are available in the supplementary material.

\subsection{Evaluation Metrics}
\label{sec:method-metrics}

For each valid (parseable) LLM output $x^{\text{LLM}}$ against reference $x^{\text{ref}}$,
we compute the relative error
\begin{equation}
  \erel = \frac{|x^{\text{LLM}} - x^{\text{ref}}|}{|x^{\text{ref}}|}
  \label{eq:rel_err}
\end{equation}
and bin cases into five categories: $<$1\%, 1--5\%, 5--10\%, 10--20\%, and $>$20\%.
The mean absolute error is
\begin{equation}
  \MAE = \frac{1}{N_{\text{valid}}}\sum_{i=1}^{N_{\text{valid}}}
         |x_i^{\text{LLM}} - x_i^{\text{ref}}|,
  \label{eq:mae}
\end{equation}
where $N_{\text{valid}}$ excludes non-parseable outputs (\textit{failures}).
Failures are counted in the $>$20\% bin to reflect end-to-end reliability;
this means MAE (computed only over parseable outputs) can appear more optimistic
than the full-distribution error picture for configurations with many failures.
$\pg$ $\MAE$ is in MW; $\qg$ $\MAE$ is in MVar.
The combined metric (\textit{All} $\MAE$, $n=50$) pools both quantities.
Because $\pg$ and $\qg$ are both power quantities on comparable numerical scales
for this network, the pooled scalar provides a useful aggregate;
separate $\pg$ and $\qg$ $\MAE$ columns are reported for completeness.
Mean relative error per configuration is reported alongside MAE as a unitless
cross-quantity metric.
No configuration tested achieves reliable accuracy sufficient for direct use as
a numerical solver; we use $\erel < 1\%$ as the threshold for \textit{solver-grade}
accuracy throughout.

\section{Results and Analysis}
\label{sec:results}

\subsection{Overall Accuracy and Prompt Format Effect}
\label{sec:results-overall}

Table~\ref{tab:full_results} reports complete results for all 12 configurations.
Fig.~\ref{fig:mae_heatmap} shows the same data as a heatmap.

\begin{table*}[!t]
\centering
\caption{Complete evaluation results for all 12 LLM--prompt configurations on 50
three-bus Gauss--Seidel power flow test cases (best All~$\MAE$ in \textbf{bold}).}
\label{tab:full_results}
\begin{tabular}{llccccccc}
\toprule
Model & Prompt Format &
All $\MAE$ & $<$5\% (\%) & $>$20\% (\%) & Fail &
$P_g$ $\MAE$ & $Q_g$ $\MAE$ \\
\midrule
Claude Sonnet 4.5 & \textit{narr.-final} & 0.772 & 38 & 54 & 0 & 0.726 & 0.836 \\
 & \textit{narr.+example} & 0.672 & 38 & 50 & 0 & 0.569 & 0.814 \\
 & \textit{proc.-text} & 0.545 & 38 & 54 & 0 & 0.199 & 1.023 \\
 & \textit{struct.-json} & 0.590 & 38 & 54 & 0 & 0.396 & 0.857 \\
\midrule
Gemini 2.5 Pro & \textit{narr.-final} & \textbf{0.257} & 54 & 22 & 1 & 0.195 & 0.346 \\
 & \textit{narr.+example} & 0.445 & 40 & 32 & 0 & 0.381 & 0.534 \\
 & \textit{proc.-text} & 0.616 & 16 & 40 & 0 & 0.689 & 0.515 \\
 & \textit{struct.-json} & 0.789 & 22 & 54 & 0 & 0.663 & 0.962 \\
\midrule
GPT-3.5 Turbo & \textit{narr.-final} & 1.820 & 2 & 90 & 0 & 2.140 & 1.379 \\
 & \textit{narr.+example} & 2.135 & 0 & 94 & 1 & 2.288 & 1.931 \\
 & \textit{proc.-text} & 2.877 & 0 & 96 & 0 & 2.940 & 2.790 \\
 & \textit{struct.-json} & 2.885 & 0 & 96 & 0 & 3.740 & 1.704 \\
\bottomrule
\end{tabular}
\end{table*}

Gemini~2.5 Pro exhibits the widest prompt-sensitivity: $\MAE$ ranges from
0.257 (\texttt{narr.-final}) to 0.789 (\texttt{struct.-json}), a 3.1$\times$ spread.
Under the \texttt{narr.-final} format, Gemini achieves the best overall result in the
benchmark: $\MAE = 0.257$~MW/MVar and 54\% of cases within 5\% relative error.
Switching to \texttt{proc.-text} more than doubles the error to 0.616, and
\texttt{struct.-json} raises it further to 0.789.

Claude Sonnet~4.5 is markedly less sensitive to prompt format.
The within-5\% fraction stays at approximately 38\% across all four formats,
while $\MAE$ varies only moderately from 0.545 (\texttt{proc.-text}) to
0.772 (\texttt{narr.-final}).
Crucially, \texttt{proc.-text}---the format that performs worst for Gemini---is the best
format for Claude, suggesting that the optimal prompt representation is model-dependent.

GPT-3.5 Turbo fails on at least 90\% of cases under all four formats,
with $\MAE$ ranging from 1.820 to 2.885.
The format that works best for Gemini (\texttt{narr.-final}) also produces the lowest
GPT-3.5 Turbo $\MAE$ (1.820 vs.\ 2.885 for \texttt{struct.-json}),
but the absolute errors remain far above practical tolerances.

\begin{figure}[!t]
  \centering
  \includegraphics[width=\columnwidth]{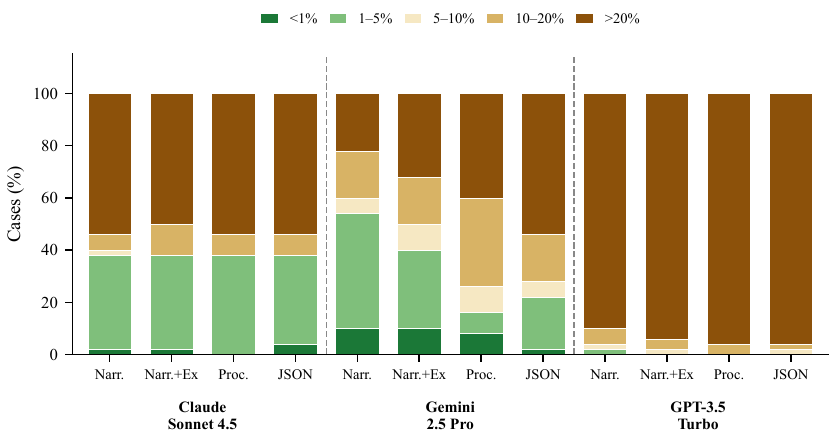}
  \caption{Distribution of case-level relative errors for all 12 model--prompt
    configurations, grouped into five bins ($n=50$; failures counted in the $>$20\% bin).}
  \label{fig:error_dist}
\end{figure}

Fig.~\ref{fig:error_dist} shows the full error distribution.
For Gemini~2.5 Pro with \texttt{narr.-final}, over half of all cases land in the
$<$5\% bins (10\% within 1\% and 44\% within 1--5\%).
Switching to \texttt{proc.-text} or \texttt{struct.-json} shifts the distribution
visibly toward the 10--20\% and $>$20\% bins.
For Claude, the distribution shape is stable across formats: all four configurations
show roughly the same $\sim$38\% within 5\% and $\sim$54\% above 20\%, consistent
with the flat $\MAE$ pattern.

\subsection{Model Capability and Prompt Sensitivity}
\label{sec:results-capability}

The three models span a wide capability range, and how much prompt format matters
tracks that range in a consistent pattern.

Adding a single solved textbook example (\texttt{narr.+example} vs.\ \texttt{narr.-final})
pulls the two capable models in opposite directions.
For Gemini~2.5 Pro the $\MAE$ rises from 0.257 to 0.445 (73\%); for Claude Sonnet~4.5
it falls from 0.772 to 0.672 (13\%), while the within-5\% fraction holds at 38\% in
both conditions.
A plausible reading of Gemini's degradation is that the worked example anchors the
model to the numerical magnitudes of that specific case, yielding miscalibrated
estimates when the target case differs substantially.
That the direction reverses for Claude rules out any single common mechanism.

At the other end of the capability spectrum, GPT-3.5 Turbo places 90--96\% of cases
above 20\% relative error regardless of prompt format.
The format effect ($\MAE$ range 1.820--2.885) is detectable but small relative to
the gap separating it from the other two models---prompt engineering alone cannot
bridge a deficit of that magnitude.

Taken together, the observed prompt-sensitivity ordering is
Gemini~2.5 Pro $\gg$ Claude Sonnet~4.5 $>$ GPT-3.5 Turbo.
The most capable model gains the most from the right prompt and loses the most from
the wrong one; the least capable model is largely unaffected because errors dominate
regardless of format.
This is consistent with prompt format modulating accuracy primarily in a middle
capability band, where the model can in principle execute the task but remains
sensitive to how instructions are presented.
The three-bus system is intentionally small, but that simplicity serves the benchmark:
it provides a clean testbed for separating model capability from task complexity, a
prerequisite for understanding LLM suitability in larger smart-grid decision-support
roles.

\subsection{Real Power vs.\ Reactive Power Accuracy}
\label{sec:results-pq}

Separating results by output type reveals a consistent pattern: $\qg$ (reactive power)
errors exceed $\pg$ (real power) errors for most model--format combinations
(Table~\ref{tab:full_results}, last two columns).
The gap is largest for Claude Sonnet~4.5 with \texttt{proc.-text}: $\pg$ $\MAE = 0.199$
while $\qg$ $\MAE = 1.023$---100\% of reactive power cases exceed 20\% relative error.
For Gemini~2.5 Pro with \texttt{narr.-final}, the gap is smaller but still present:
$\pg$ $\MAE = 0.195$ vs.\ $\qg$ $\MAE = 0.346$.

This asymmetry aligns with the computational structure of the problem.
Computing $\pg$ requires extracting the real part of $\bar{S}_1$,
while $\qg$ requires the imaginary part of the same complex product.
Both require forming $\bar{S}_1 = \bar{V}_1 \cdot \overline{\sum_j Y_{1j}\bar{V}_j}$,
but models appear to handle the imaginary extraction with lower fidelity.
One hypothesis is that real-part extraction is more heavily represented in power
systems training text (textbook problems more often request real power), making
imaginary-part extraction a less practiced operation; this remains unverified.

\subsection{Generalizability: 100-Case Replication}
\label{sec:results-scale}

\begin{figure}[t]
  \centering
  \includegraphics[width=\columnwidth]{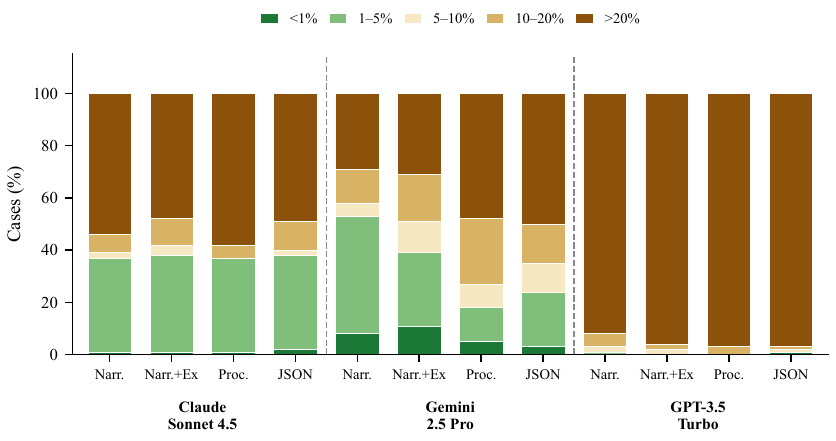}
  \caption{Error-bin distribution for the 100-case replication ($n=100$;
    failures counted in the $>$20\% bin; format labels correspond to
    Table~\ref{tab:results_100}).}
  \label{fig:error_dist_100}
\end{figure}

Table~\ref{tab:results_100} and Fig.~\ref{fig:error_dist_100} report results for
100~independently drawn test cases (same parameter ranges; see
Section~\ref{sec:method-system}) under four related prompt-format families.

Three qualitative conclusions from the 50-case benchmark replicate without exception.
The model capability ordering (Gemini\,$\gg$\,Claude\,$>$\,GPT-3.5) is preserved.
GPT-3.5 Turbo places 92--97\% of cases above 20\% relative error under all formats,
consistent with 90--96\% at 50~cases.
No configuration achieves solver-grade accuracy.

Claude Sonnet~4.5 again exhibits a near-flat format sensitivity profile: the
within-5\% fraction holds at 37--38\% and $\MAE$ spans 0.515--0.812 across all
four formats, confirming that format insensitivity is a robust model property
rather than a sampling artifact.

Gemini~2.5 Pro's best format shifts from \texttt{narr.-final} (50~cases) to
\texttt{proc.-text} at 100~cases (MAE\,=\,0.402, 53\% within 5\%), while
\texttt{narr.-final} rises to MAE\,=\,0.700.
This format-ranking reversal is consistent with higher task difficulty or sampling
variability; single-run evaluation cannot distinguish these explanations.
Notably, GPT-3.5 Turbo also achieves its lowest 100-case $\MAE$ under
\texttt{proc.-text} (1.834 vs.\ 2.866 for \texttt{narr.-final}), suggesting that
explicit step-by-step output demands provide a weak benefit even at low
capability---though 92\% of cases still exceed 20\% error.

\begin{table}[t]
\centering
\caption{100-case replication results (best All~$\MAE$ in \textbf{bold};
  $^\dagger$2~non-parseable outputs included in $>$20\% bin).}
\label{tab:results_100}
\begin{tabular}{llccc}
\toprule
Model & Format & All $\MAE$ & $<$5\% (\%) & $>$20\% (\%) \\
\midrule
Claude     & \textit{narr.-final}   & 0.524 & 37 & 58 \\
Sonnet 4.5 & \textit{struct.-json}  & 0.515 & 38 & 49 \\
           & \textit{proc.-text}    & 0.812 & 37 & 54 \\
           & \textit{narr.+example} & 0.623 & 38 & 48 \\
\midrule
Gemini     & \textit{narr.-final}   & 0.700 & 18 & 48 \\
2.5 Pro    & \textit{struct.-json}  & 0.651 & 24 & 50 \\
           & \textit{proc.-text}    & \textbf{0.402} & 53 & 29 \\
           & \textit{narr.+example} & 0.417 & 39 & 31 \\
\midrule
GPT-3.5    & \textit{narr.-final}   & 2.866 &  0 & 97 \\
Turbo      & \textit{struct.-json}  & 3.292 &  1 & 97 \\
           & \textit{proc.-text}    & 1.834 &  1 & 92 \\
           & \textit{narr.+example} & 2.191$^\dagger$ & 0 & 96 \\
\bottomrule
\end{tabular}
\end{table}

\section{Conclusion}
\label{sec:conclusion}

We benchmarked three LLMs across four prompt formats on Gauss--Seidel AC power flow
for 50 three-bus test cases.
The main finding is that increasing prompt complexity---combining structured
data representation with required iteration-trace output---correlates with
significantly lower accuracy for Gemini~2.5 Pro (3.1$\times$ MAE increase)
and does not benefit Claude Sonnet~4.5.
Adding a worked example further degrades accuracy for Gemini.
The optimal prompt format differs by model, and no configuration tested achieves
reliable accuracy sufficient for direct use as a numerical solver.

These results have two practical implications.
First, practitioners deploying LLMs for AC power flow assistance should not assume that
more detailed prompting always helps---the effect is model-specific and can be
counterproductive.
Second, none of the tested models is ready to replace a numerical solver; they may
be useful as approximate reasoning aids or for educational explanation, but not for
production computation.

An independent 100-case replication confirms that the capability ordering and the
no-solver-grade-accuracy conclusion are robust to sample scale.
Claude's format insensitivity replicates exactly; Gemini's optimal format shifts
from \texttt{narr.-final} to \texttt{proc.-text} at 100~cases, suggesting that
prompt-format sensitivity may be modulated by instance difficulty---a direction
for future investigation under controlled difficulty scaling.

\bibliographystyle{IEEEtran}
\bibliography{references}

\end{document}